# Artificial Intelligence-assisted Pixel-level Lung (APL) Scoring for Fast and Accurate Quantification in Ultra-short Echo-time MRI

Bowen Xin[1#], Rohan Hickey[2], Tamara Blake[2,3], Jin Jin[4], Claire E Wainwright[2,3], Thomas Benkert[5], Alto Stemmer[5], Peter Sly[2,3], David Coman[2,6], Jason Dowling[1]

[1] Australian E-health Research Centre, CSIRO, Brisbane, Queensland, Australia

[2] School of Medicine, The University of Queensland, Brisbane, Queensland, Australia

[3] Department of Respiratory and Sleep Medicine, Queensland Children's Hospital, Brisbane, Queensland, Australia

[4] Siemens Healthcare Pty Ltd, Brisbane, Queensland, Australia

[5] Application Predevelopment, Siemens Healthineers AG, Forchheim, Germany

[6] Queensland Lifespan Metabolic Medicine Service, Queensland Children's Hospital, Brisbane, Queensland, Australia

## INTRODUCTION

Lung magnetic resonance imaging (MRI) with ultrashort echo-time (UTE) represents a recent breakthrough in lung structure imaging, providing image resolution and quality comparable to computed tomography (CT)[1]. Due to the absence of ionising radiation, MRI is often preferred over CT in paediatric diseases such as cystic fibrosis (CF), one of the most common genetic disorders in Caucasians[2,3]. To assess structural lung damage in CF imaging, CT scoring systems provide valuable quantitative insights for disease diagnosis and progression[4]. However, few quantitative scoring systems are available in structural lung MRI (e.g., UTE-MRI).

Specifically, early semi-quantitative lung MRI scoring systems[5,6] were not designed to sensitively assess the subtle structural pulmonary changes. Lately, a quantitative grid-based scoring system[7] was adapted for lung MRI from PRAGMA-CF scoring in CT[4]. However, this scoring system requires time-consuming manual lung annotation, and this grid-based system does not provide pixel-level annotation accuracy. To provide fast and accurate quantification in lung MRI, we investigated the feasibility of novel Artificial intelligence-assisted Pixel-level Lung (APL) scoring for CF.

## METHODS

Data acquisition: Data were acquired with a 3D UTE stack-of-spirals VIBE research application sequence [8,9] on a 3T scanner (MAGNETOM Prisma; Siemens Healthineers, Forchheim, Germany). Fourteen patients with cystic fibrosis and ten healthy volunteers were scanned, following local ethics approval, with all participants providing informed consent. Combined prospective and retrospective gating based on a superoinferior readout self-navigator signal was used. Scans were performed with free breathing. Imaging parameters: acquisition time 10-15 min, resolution 1.1 or 1.5 mm isotropic, FOV = 544*544 or 480*480 mm, TE/TR = 3.78/0.05 ms, Flip angle = 5˚.

APL scoring: As described in Figure 1, APL scoring consists of 5 stages, including 1) image loading, 2) AI lung segmentation, 3) lung-bounded slice sampling, 4) pixel-level annotation, and 5) quantification and reporting. The scoring process has been highly automated except for pixel-level annotation. Specifically, after image loading, the left and right lungs are automatically segmented using a developed APL-seg deep learning model (please see next paragraph) to replace manual annotation. Within the lung boundary, the system uniformly samples ten axial slices for the subsequent pixel-level annotation of bronchiectasis/ airway thickening, mucus plugging, and consolidation/atelectasis. Lastly, the ratio of diseased volume over sampled lung volume (diseased-to-lung ratio) is calculated and reported. APL scoring was tested on the 14 UTE-MRI of CF patients by an imaging scientist with 3 years of experience in lung MRI (BX). The software was developed as a plug-in in the 3D Slicer platform [10] using Python 3.9.

AI Lung segmentation (APL-seg): We locally trained three lung segmentation networks, including a 2D model (APL-seg2d), a 3D low-resolution model (APL-seg3d-L), and a 3D full-resolution model (APL-seg3d-F) using the nnUNet-v2 framework [11]. 24 UTE-MRIs were randomly divided, with 20 used for training and 4 for independent testing. We compared our trained models with publicly available MRI lung segmentation algorithms, including Antspynet [12] and TotalSegmentator [13], and evaluated the performance using the Dice score. All the testing was performed on a Nvidia H100 GPU.

Statistical analysis: We conducted a paired $t$-test to analyse the time efficiency and annotation accuracy between APL and quantitative grid-level scoring [7]. Pearson's correlation coefficient was used to test the consistency between the two scores. A two-tailed $p$-value below 0.05 was considered statistically significant.

## RESULTS

Comparison of APL and grid-level scoring: A visual comparison of APL scoring and conventional grid-based scoring was illustrated in Figure 2. Figure 3(a) shows that our APL scoring took 8.2 minutes per subject, which was more than twice as fast as the previous grid-level scoring (17.5 minutes per subject, $p$=1.48e-18). Figure 3(b) shows our pixel-level scoring was statistically more accurate ($p$=0.021), while Figure 3(c) shows APL scoring was still strongly correlated with grid-level scoring (R=0.973, $p$=5.85e-9).

<u>Comparison of AI lung segmentation:</u> Figure 4 show that our three APL-seg models (Dice 0.969-0.974) consistently outperformed Antspynet (Dice 0.878) and TotalSegmentator (Dice 0.955). In Figure 5, our APL-seg3d-F model demonstrates superior segmentation results visually, particularly in the central lung region.

**DISCUSSION AND CONCLUSION**

In this work, AI-assisted Pixel-level Lung (APL) scoring achieved efficient and accurate outcomes in UTE lung MRI. The APL scoring process is predominantly driven by automated AI lung segmentation, which substantially reduces the scoring time and facilitates savings on the time needed for pixel-level accurate annotation. This tool has great potential to streamline the workflow of UTE lung MRI in clinical settings, and be extended to other structural lung MRI sequences (e.g., BLADE MRI), and for other lung diseases (e.g., bronchopulmonary dysplasia).

## Acknowledgement

We acknowledge the contribution of students from the University of Sydney (Lachlan Muir, Matthew Butler, Guorui Jin, Luca Lamond, Hugo Taranto, Tuan Minh Nguyen, and Will Neuner) to the development of the software.

**Figures**

Figure 1. Workflow of AI-assisted Pixel-level Lung (APL) scoring for ultrashort echo-time (UTE) MRI

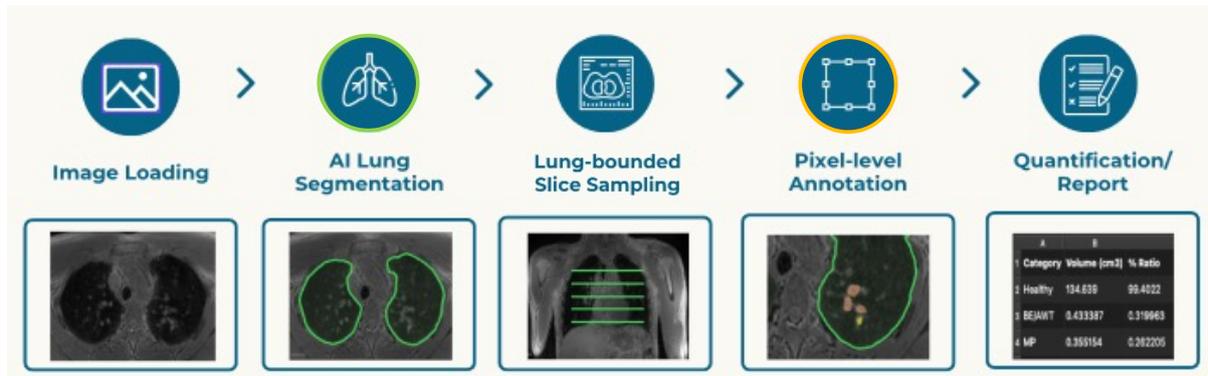

Figure 2. Visual comparison of (a) conventional grid-level scoring and (b) our AI-assisted Pixel-level Lung scoring. The scoring was performed on a 22-year-old female CF patient. The red colour indicates bronchiectasis/ airway thickening. The yellow colour indicates the mucus plug.

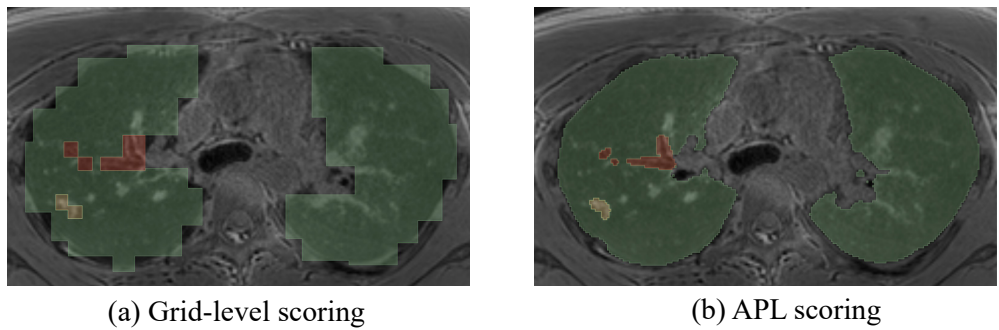

(a) Grid-level scoring                      (b) APL scoring

Figure 3. Quantitative comparison of APL and grid-level scoring regarding (a) time efficiency and (b) scoring results, and (3) correlation analysis of two scores. Our APL scoring is statistically faster and more accurate than grid-level scoring, while strongly correlated with the grid-level scoring. * denotes $p<0.05$, ** denotes $p<0.01$, *** denotes $p<0.001$, and **** denotes $p<0.0001$.

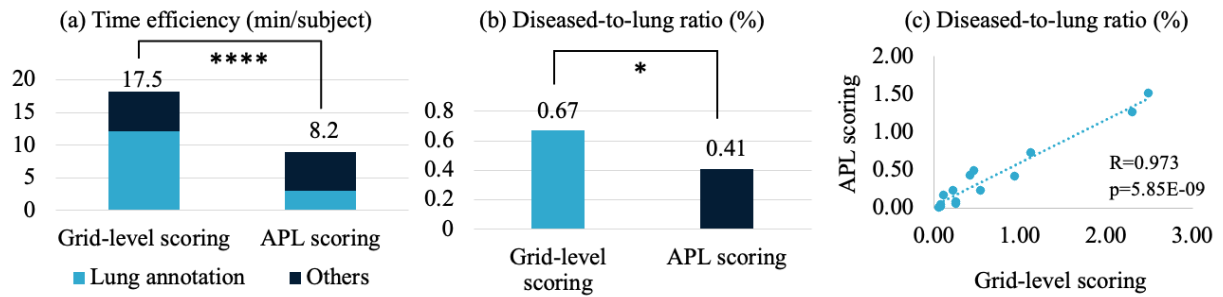

Figure 4. Quantitative results for automated UTE-MRI lung segmentation on the test set. The red border indicates the most accurate segmentation model.

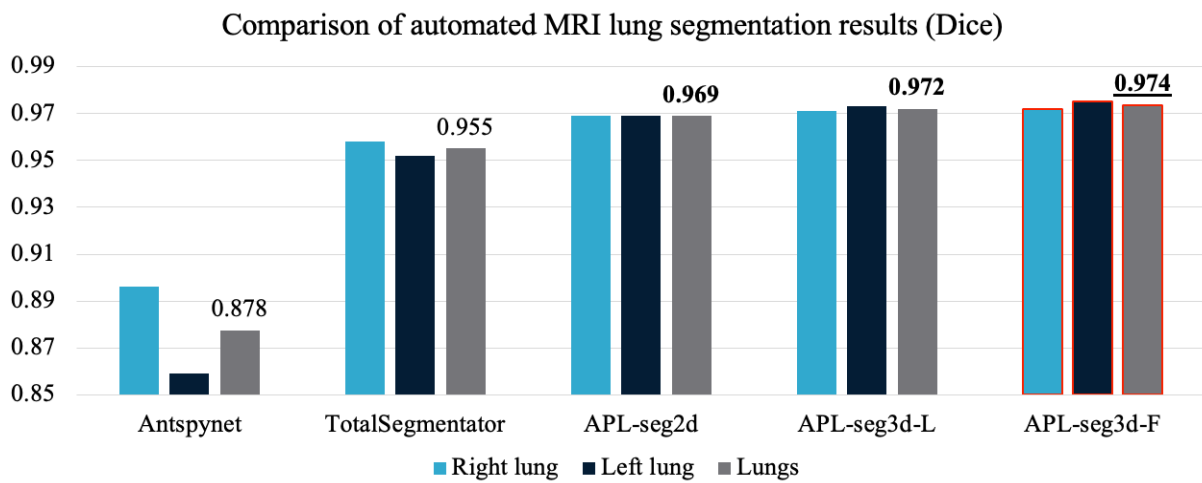

Figure 5. Qualitative comparison for automated UTE-MRI lung segmentation. The green and yellow area indicates the segmented lung volumes, while the red contour indicates the ground truth. The red arrows highlight that our APL-seg3d-F model demonstrated superior segmentation performance in the central lung region.

(a) Antspynet

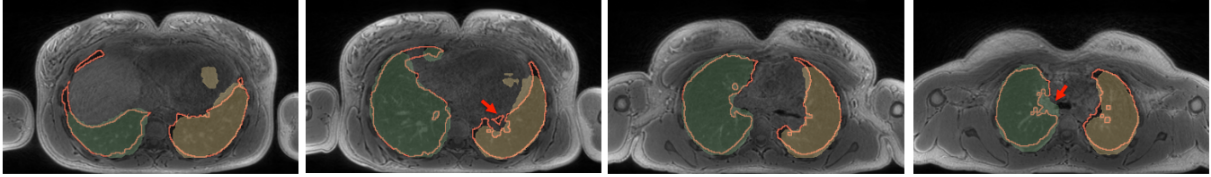

(b) TotalSegmentator

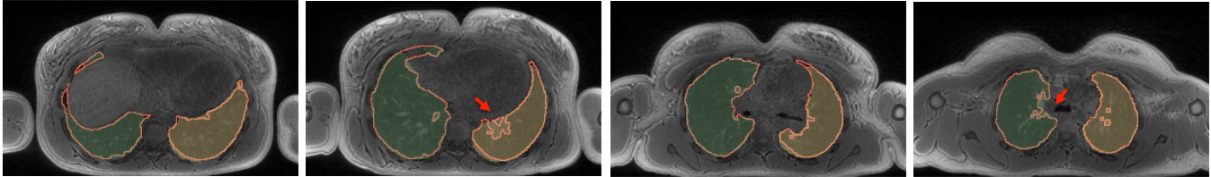

(c) APL-seg3d-F

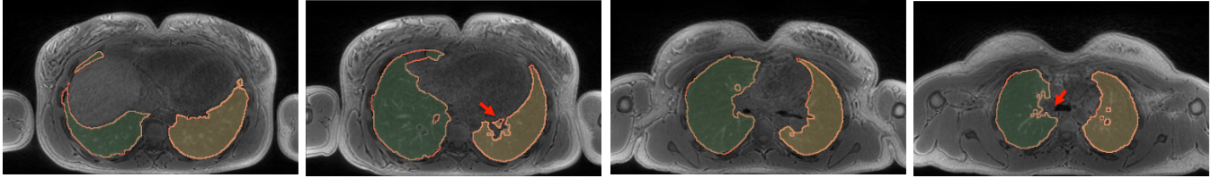